\documentstyle[12pt]{article}

\newcommand{\be}{\begin{equation}}
\newcommand{\ee}{\end{equation}}
\newcommand{\ba}{\begin{eqnarray}}
\newcommand{\ea}{\end{eqnarray}}

\begin{document}
\hoffset=-.4truein\voffset=-0.5truein
\setlength{\textheight}{8.5 in}
\begin{titlepage}
\begin{center}
\vskip 16 mm

{\large \bf Punctures and $p$-spin curves from matrix models II }
\vskip .6 in
\begin{center}
{\bf S. Hikami$^{a)}$}{\it and} {\bf E. Br\'ezin$^{b)}$}
\end{center}
\vskip 5mm
\begin{center}

{$^{a)}$ Okinawa Institute of Science and Technology Graduate University, 1919-1 Tancha, Okinawa 904-0495, Japan.
e-mail: hikami@oist.jp
} \\ {$^{b)}$ Laboratoire de Physique de l'Ecole normale sup\'erieure, ENS,
Universit\'e PSL, CNRS, Sorbonne Universit\'e, Universit\'e de Paris, F-75005 Paris, e-mail: brezin@lpt.ens.fr} \\
\end{center}

\vskip 3mm       
{\bf Abstract}                  
\end{center}
   We report here an extension of a previous work in which we have shown that matrix models provide a tool to compute the intersection numbers of  $p$-spin curves. 
   We discuss   further  an extension to half-integer $p$, and in more details for $p=\frac{1}{2}$ and $p=\frac{3}{2}$. In those new cases one finds contributions
   from the Ramond sector, which were not present for positive integer $p$.
   The existence of Virasoro constraints, in particular a string equation, is considered also for half-integral spins.
   The contribution of the boundary of a Riemann surface,  is investigated through  a logarithmic matrix model.
   The supersymmetric random matrices provide extensions to mixed positive and negative $p$ punctures.
   
\end{titlepage}

\vskip 3mm

 \section{Introduction}

We present here a continuation of a previous article \cite{BrezinHikami2020}, where the extension to half-integer  $p$ spins was introduced on the basis of explicit expressions for   random matrix
 correlation functions \cite{BrezinHikami2}. The marked points on Riemann surfaces of $p$-spin curves are classified into two types, Neveu-Schwarz (NS) and Ramond (R). In this paper, we investigate the correlation functions up to three marked points.

 The integer $p$-spin curves are the moduli spaces of Riemann surface with spin structures and they are described by matrix models, in which $p$  appears as a power  of the random  matrices.
 In a previous article \cite{BrezinHikami03}, the values of $p$ were continued to  negative values such as $p=-1, -2$, where $p=-1$ corresponds to
 Euler characteristics.  The  $p=-2$ spin case corresponds to a unitary matrix model, which is simply the one-plaquette model of a lattice  gauge theory.  This well-known  model exhibits an interesting phase
 transition between the weak  and strong coupling regimes \cite{GrossWitten,BrezinGross}.
 
 It  is natural to examine  the continuation to half-integer $p$. 
 Starting from the integer spin expressions, in a previous article \cite{BrezinHikami2020}, we have introduced a continuation to half- integer spin cases, focusing on the typical case  $p=\frac{1}{2}$.  The continuation to $p=\frac{1}{2}$ involves the removal of an infinite renormalization  factor $\Gamma(0)$. 
 A remarkably   simple and exact result for the  one point function, for general genus $g$, is obtained for $p=\frac{1}{2}$ within this renormalization.
 
In the case of  integer $p$-spin,  the spin components take $p$ possible values;  the $(p-1)$ values  $l= 0,1,2,...,p-2$,   
correspond to  Neveu-Schwarz(NS)  marked points ; for   $l= p-1$ or  $l= -1$ (mod $p$), the marked point on the Riemann surface  is of  Ramond (R)- type. In this article, we use those definitions of NS  and R type, classified by the value of $l$ \cite{Witten1}.
 In the case of positive integer $p$, we know that  to  $p=2$ corresponds a KdV hierarchy, and a  generalized KdV for $p\ge 3$. For $p=\frac{1}{2}$, the existence of a related integrable
structure is  not known.

A  selection rule implies that the R-type punctures should appear in  pairs for $p=-2$, and we show explicitly the occurrence of pair-wise punctures for the three-point function.

The organization of this article is the following: The section two is devoted to spin $p=\frac{1}{2}$ and its  one-point  correlation function. We verify the  continuation of the expression for
$p$-spin curves to the $p=\frac{1}{2}$ case.
The section three deals with the two-point function for $p=\frac{1}{2}$. A  possible string equation is discussed.
 The section four gives the three-point functions for $p=\frac{1}{2}$. 
The spin $p=\frac{3}{2}$ is studied in  the section five.
  In section six, the negative
integer cases  $p=-2$ and $p=-3$ are investigated. We discuss the paired R-type punctures for $p=-2$. 
In section seven, we discuss  supermatrices and their correlation functions. The strong coupling expansions are considered. 
We discuss the open boundary and  marked points on the boundary within a  logarithmic matrix model. Through this open boundary analysis, we find the 
 continuation from integer $p$ to the $p=\frac{1}{2}$ case.
 The  section eight contains a  summary and a discussion.

 \section{Spin $p = \frac{1}{2}$ correlation functions}
 \vskip 3mm
 There are two types of marked points on Riemann surfaces : Neveu-Schwarz(NS) and Ramond (R).  
 In the spin $p=\frac{1}{2}$ case, the marked points on the Riemann surface belong to the Ramond type  with $l=p-1= -\frac{1}{2}$, or $l=0,-1$ (modulo $p$, these three cases are in fact all the same one). The choice of $l$ among these three cases depends upon the 
 continuation from  integer $p$ to half-integer $p$ (see the discussion below).
 
 The intersection numbers for positive integer $p$ are given by the integral of the compact moduli space $\bar M_{g,s}$,
 \be\label{intersection number}
 <\tau_{n_1,l_1}\cdots \tau_{n_s,l_s}>_g = \frac{1}{p^g} \int_{{\bar M}_{g,s}} {c_1({\mathcal{L}}_1)}^{n_1} \cdots { c_1({\mathcal{L}}_s) }^{n_s} C_D(V)
 \ee
 where $c_1({\mathcal{L}}_j)$ is the first Chern class 
 and $C_D(V)$ is a top Chern class
 with dimension $D= (g-1)(1- \frac{2}{p}) + \frac{1}{p}\sum l_i$. 
 
 In this work we continue those formulae to half-integer $p$.
 In previous articles, we have found that the correlation functions of a random-matrix theory provide  intersection numbers for the moduli of curves on  a Riemann surface.
More precisely, the intersection numbers  (\ref{intersection number}) are deduced from  the correlation functions $U(\sigma_1,....,\sigma_s)$ \cite{BrezinHikami03,BrezinHikami05,BrezinHikami01},
 \be\label{Us}
 U(\sigma_1,...,\sigma_s) = < \rm{tr} e^{\sigma_1 M} \cdots \rm{tr} e^{\sigma_s M} >
 \ee
 where $M$ is a Gaussian random matrix coupled to an appropriately tuned matrix source $A$.
 Those correlation functions are known to satisfy  explicit integral representations \cite{BrezinHikami2}.
 For the spin $p$ case, after  tuning of the external source $A$, it reduces to
 \be \label{s-point}
 U(\sigma_1,...,\sigma_s) = \oint \prod_{i=1}^s \frac{du_i}{2\pi i} e^{C [(u_i - \frac{\sigma_i}{2})^{p+1} - (u +  \frac{\sigma_i}{2})^{p+1}]}
 \prod_{i<j} \frac{1}{u_i-u_j + \frac{1}{2}(\sigma_i+ \sigma_j)}
 \ee
 The intersection numbers (\ref{intersection number}) are obtained as the coefficients of the expansion in powers of the $\sigma_i$'s in this matrix model approach, with appropriate normalizations  \cite{BrezinHikami2},
  \be
 U(\sigma_1,...,\sigma_s) = \sum_{n_i,l_i} \prod_{i=1}^s \sigma^{n_i+ \frac{1}{p}(1+ l_i)} <\tau_{n_1,l_1}\cdots \tau_{n_s,l_s}>
 \ee

 The non-vanishing the intersection numbers are restricted by a selection rule for the  parameters $(n_i,l_i)$ 
 \be\label{RR}
 3 g -3 + s = \sum_{i=1}^s n_i + ( g-1) ( 1 - \frac{2}{p}) + \frac{1}{p}\sum_{i=1}^s l_i
 \ee

 The second and third term is denoted as $D$, which is derived from Riemann-Roch formula \cite{Witten1}.
 This formula can be written as
\be
2g-2 + s = \frac{1}{p+1}\sum_{i=1}^s (1 + l_i + p n_i)
 \ee
 where the left hand side is minus Euler characteristics $- \chi $.
 
When $p= \frac{1}{2}$, where all $l_i= l$, and above equation is written as 
\be\label{chi}
6g - 6 + (1- 2l)  s  = \sum_{i=1}^s n_i
\ee
where $n_i$ is the exponent of ${\sigma_i}^{n_i + 2(1+ l_i)}$ for $U(\sigma_1,...,\sigma_s) \sim \prod_{i=1}^s \sigma^{n_i+ 2(1+ l_i)}$.

 We consider  the one-point function $ <{\rm tr} e^{\sigma M} >$ in the scaling limit \cite{BrezinHikami2020} :  
\ba\label{uintegral}
U(\sigma) &=& \frac{1}{\sigma}\oint \frac{du}{2i \pi} e^{\frac{\xi}{p+1}((u+ \frac{1}{2}\sigma)^{p+1} - (u - \frac{1}{2}\sigma)^{p+1})}\nonumber\\
&=&\oint \frac{du}{2i \pi} e^{ c \sigma^{p+1}((u+ \frac{1}{2})^{p+1} - (u - \frac{1}{2})^{p+1})}
\ea
where we  have rescaled $u \to \sigma u$ , and $c= \xi/(p+1)$.
We now change $u$ to a  variable $y$  better suited to discuss half-integer $p$ \cite{BrezinHikami2020}
\be\label{zy}
u = \frac{i}{2} ( y^2- y^{-2} ), \hskip 3mm du = i ( \frac{1}{y^3} + y ) dy.
\ee

The one point function $U(\sigma)$ of (\ref{uintegral}) becomes
\be\label{double}
U(\sigma)= \frac{i}{2}\oint \frac{dy}{2i\pi} (y+ \frac{1}{y^3})e^{c(\frac{i \sigma}{4})^{p+1} \frac{1}{y^{2p+2}}[(y^2-i)^{2p+2}-(y^2+i)^{2p+2}]}
\ee

 This integrand, which has  singularities at $y=0$ and $y=\infty$,  leads to useful  contour integrals. It is invariant under the transformation 
 $y\to - \frac{1}{y}$. 
 
   For instance, we obtain the  one-point function,
 \be\label{onepoint1}
 U(\sigma) =\frac{i}{2\sigma} \oint \frac{dy}{2\pi i}  (y+ \frac{\sigma^2}{y^3}) g(y) 
 \ee
 where for $p= \frac{1}{2}$, $g(y)$ is given as
 \be\label{gy}
 g(y) = e^{c' \sigma ( 3 y - \frac{\sigma^2}{y^3})}
 \ee
 
Since the exponent is linear in $y$ in the large $y$ limit,  we can deform    the   contour in (\ref{onepoint1}) to a large circle. If we expand for  small $\sigma$, a singularity appears at the
origin $y=0$ or at  $y=\infty$.
From the residues at $y=0$, the one point function is given by 
\ba\label{onepoint}
 U(\sigma) &=& \frac{i}{2}\sum_{j=0}^\infty c'^{4j+2} \frac{3^{2+3j}}{j!(3j+3)!}(-1)^j \sigma^{3+ 6 j}\nonumber\\
 &=& \frac{i}{2}\sum_{g=1}^\infty c'^{4g-2} \frac{3^{3g-1}}{(g-1)!(3g)!}(-1)^{g-1} \sigma^{6g-3}
 \ea
 Using  the  Gauss-Legendre formula for $\Gamma(3 z)$,
  $
  \Gamma(3z) = \frac{3^{3z}}{(2\pi)\sqrt{3}}\Gamma(z)\Gamma(z+ \frac{1}{3})\Gamma(z+ \frac{2}{3}),
  $
  the expression (\ref{onepoint}) becomes
  $
   U(\sigma)= \frac{i \pi \sqrt{3}}{9} \Gamma(\frac{4}{3})\Gamma(\frac{5}{3}) {}_0F_{3} (\frac{4}{3},\frac{5}{3},2; - c'^4 \sigma^6)
$
  where ${}_{0}F_3(\beta_1,\beta_2,\beta_3;z)$ is a standard generalized hypergeometric function.

 Integer powers  of (\ref{onepoint}) are indeed expected, since for the general $p$-spin curve, we argued that the powers  are of the form  $\sigma^{n+ \frac{1}{p}(1+ l)}$ which is an integer for $p=\frac{1}{2}$. Using the dependence in $n$ and $l$ in the general form $\sigma^{n+2(1+l)}$, we find an integral power of $\sigma$ for $l=0,-\frac{1}{2},-1$ with the selection rule $n= 6g-5 - 2l$ in (\ref{chi}).  The coefficient of $\sigma^{n+2(1+l)}= \sigma^{6g-3}$  is denoted by $<\tau_{6g-3}>$ (it is independent of the choice of $l$).  

To verify the consistency with the known expressions for   integer $p$, we return to  previous results  \cite{BrezinHikami2}.
The one-point function was computed  up to $g=9$ for general integer $p$ in \cite{BrezinHikami2}. Let us  quote here the results  up to $g=3$,
\ba\label{3/2u}
&&U(\sigma)= \frac{1}{\sigma}\int \frac{du}{2i\pi} e^{-\frac{c}{p+1}[(u+\frac{\sigma}{2})^{p+1}-(u-\frac{\sigma}{2})^{p+1}]}\nonumber\\
&&=\frac{1}{\sigma p \pi} \cdot \frac{1}{(cp)^{1/p}} \int_0^\infty dt t^{\frac{1}{p}-1}e^{-t}
 e^{-\frac{p(p-1)}{3! 4}\sigma^{2+ \frac{2}{p}}c^{\frac{2}{p}} t^{1-\frac{1}{p}}- \frac{p(p-1)(p-2)(p-3)}{5! 4^2}\sigma^{4+\frac{4}{p}}c^{\frac{4}{p}}t^{1-\frac{4}{p}}+ \cdots}\nonumber\\
 \nonumber\\
&&=\frac{1}{\sigma\pi}\cdot \frac{1}{(c\sigma)^{1/p}}\biggl(\Gamma(1+ \frac{1}{p})-\frac{p-1}{24} z \Gamma(1-\frac{1}{p})\nonumber\\
&&+ \frac{(p-1)(p-3)(1+ 2 p)}{5!\cdot4^2\cdot 3} z^2 \Gamma(1-\frac{3}{p}) \nonumber\\
&& - \frac{(p-1)(p-5)(1+ 2 p) (8p^2-13 p - 13)}{7!4^33^2} z^3 \Gamma(1 - \frac{5}{p}) \nonumber\\
&&+  \frac{(p-1)(p-7)(1+ 2p) (72 p^4 -298 p^3 -17 p^2+562 p+281)}{9!4^4 15}z^4 \Gamma(1-\frac{7}{p})\nonumber\\
&& + O(z^4)\biggr)
\ea
where $z=c^{\frac{2}{p}}\sigma^{2+ \frac{2}{p}}$.
Performing the continuation to  $p=\frac{1}{2}$ in the above formula, we obtain divergent results with coefficients $\Gamma(-n)$. The limit $p\to 1/2$ should thus be taken with care. If we  replace $\Gamma(-n +\epsilon)$ by $ (-1)^n \frac{1}{n!}\Gamma(\epsilon)$, where $\epsilon$ is proportional to $p-1/2$  we obtain the divergent expression
\ba\label{exact2}
U(\sigma) &=& [ \frac{1}{48}c^2\sigma^3 + \frac{1}{6! 4^3\cdot 6} c^6\sigma^9 + \frac{1}{9! 3^2 4^6} c^{10}\sigma^{15}
+ \frac{1}{(12)! 4^8 3^4} c^{14}\sigma^{21} + \cdots ] \Gamma(\epsilon)\nonumber\\
\ea

Nevertheless this expression may be compared to $U(\sigma)$ of $p=\frac{1}{2}$ with $c' = \frac{1}{6}c$ in (\ref{onepoint}), 
\be\label{pint2}
U(\sigma) = i [ \frac{1}{48}c^2\sigma^3 + \frac{1}{6! 4^3\cdot 6} c^6\sigma^9 + \frac{1}{9! 3^2 4^6} c^{10}\sigma^{15}
+ \frac{1}{(12)! 4^8 3^4} c^{14}\sigma^{21} + \cdots ]
\ee
Thus except for a factor $i\Gamma(0)$, both expressions coincide. The divergent factor  $\Gamma(0)$ is due to the integral (\ref{3/2u}), whereas
for $p=\frac{1}{2}$, we had used the contour integral for $y$ of (\ref{onepoint1}). Thus, modulo this  infinite renormalization, we have found  a remarkable extension to $p=\frac{1}{2}$.
Note that  in (\ref{3/2u}) the the gamma factors are also divergent for  integer $p$, but these divergences are  cancelled by the associated pre-factors.

To  clarifiy  the  question  of divergences related to R-punctures, we  return here to earlier  results for $U(\sigma)$. As shown  in \cite{BrezinHikami2}, the generating function of the one-point  intersection numbers $U(\sigma)$ is known in closed form  for $p=2,3,4$.
 For $p=2$, it is given by
 \be
 U(\sigma) = \frac{1}{2\sqrt{\pi}\sigma^{\frac{3}{2}}} e^{\frac{\sigma^3}{12N^2}}
 \ee
 which shows that there is no integer power in the expansion in powers of $\sigma$.
 For $p=3$ , we have \cite{BrezinHikami09}
 \ba\label{p3}
 U(\sigma) &=& \frac{1}{N\sigma(3N\sigma)^{1/3}} A_i \biggl( - \frac{N^{2/3}}{4\cdot 3^{1/3}}\sigma^{8/3}\biggr)\nonumber\\
 &=& 
 \frac{1}{N\sigma(3N \sigma)^{1/3}}\biggl[ A_i(0) \biggl( 1 + \frac{1}{3!}\zeta^3 +\frac{4}{6!}\zeta^6 + \cdots\biggr)
 + A'_{i}(0)\biggl( \zeta + \frac{2}{4!}\zeta^4 \cdots\biggr)\biggr]\nonumber\\
 \ea
where $\zeta= - N^{2/3}(4\cdot 3^{1/3})^{-1}\sigma^{8/3}$, and $A_i(0)= 3^{-2/3}/\Gamma(2/3)$, $A'_i(0) = - 3^{-1/3}/\Gamma(\frac{2}{3})$.
The asymptotic expansion of the Airy function consists of two series. The first series is related to $l=1$ and the second series related to $l=0$; they provide
\be
<\tau_{1,0}>_{g=1} = \frac{1}{12}, \hskip 3mm <\tau_{6,1}>_{g=3} = \frac{1}{3 \cdot 3! (12)^3}, \hskip 3mm
<\tau_{9,0}>_{g=4}= \frac{1}{(12)^4 4!}\cdot \frac{2}{3}, \cdots
\ee
 There is no R-type term  of the type $<\tau_{n,2}>$ or $<\tau_{n,-1}>$, which  would appear 
 as integer powers of $\sigma$ in the expansion of $U(\sigma)$ (note $l=2$ and $l=-1$ are identical mod $p=3$).
 
For $p=4$, the one point function reads \cite{BrezinHikami2}
\ba
U(\sigma) &=& \frac{1}{2\sqrt{8}} e^{\frac{3}{160}\sigma^5} \frac{1}{2{\rm sin}(\frac{\pi}{4})} 
( I_{-\frac{1}{4}}(\frac{1}{32}\sigma^5) + I_{\frac{1}{4}}(\frac{1}{32}\sigma^5) ) \nonumber\\
&=& \frac{1}{8}\sum_{m,n=0}^\infty \frac{1}{m!n!\Gamma(n+ \frac{5}{4})} (\frac{3}{160})^m(\frac{1}{64})^{2n+ \frac{1}{4}}\sigma^{5m+10n+\frac{1}{4}}
\ea
which shows that  all terms are of NS-type (the power of $\sigma$ is not  an integer) and none of  R-type.
Returning to (\ref{3/2u}) we see that the coefficients present divergent gamma-functions of negative integers, but these divergences disappear if one takes into account the pre-factors, as for instance with $ (p-3) \Gamma(1-3/p)$ which goes to 3 for $p\to 3$.  This procedure does not provide any non-vanishing R term. 

To incorporate  R type intersection numbers, we return to the relationship between the intersection numbers $<\tau_{n,l}>$ and the generating function   $U(\sigma)$ 
\be\label{def}
U(\sigma) = \sum_g <\tau_{n,l}>_g\frac{1}{\pi}\Gamma(1 - \frac{1+l}{p}) p^{g-1}\sigma^{(2g-1)(1+ \frac{1}{p})}
\ee
where $(p+1)(2g-1)=pn+ l+1$. The intersection number $<\tau_{n,l}>_g$ has to vanish to cancel the divergent factor 
$ \Gamma(1 - \frac{1+l}{p})$ for $ l=p-1$ (R type). For $p=3$,  the
intersection numbers  are known in closed form \cite{BrezinHikami01},
\be
<\tau_{n,l}>_g = \frac{1}{(12)^g g!} \frac{\Gamma(\frac{1+g}{3})}{\Gamma(1 - \frac{1+l}{3})}
\ee
with the selection rule  $\frac{8}{3}(g-1)+ 1 = n + \frac{l}{3}$. Thus the infinite denominator implies vanishing intersection numbers of the would be R-type for $g=2+ 3m$ ($m=0,1,2,...$).

Therefore, if we make  a continuation to half-integer $p$, the divergence of $U(\sigma)$ in (\ref{exact2}), is matched by 
 the divergence of the  gamma function $\Gamma(1-\frac{1+l}{p})$ in (\ref{def}), when we take $l=p-1$ . This allows for finite non-vanishing  limits of $<\tau_{n,l}>_g$ for $l=p-1$. The definition  (\ref{def}) for the intersection numbers provides the normalization for the factorization of the divergence even for half-integer $p$. Namely one should consider the limit $p\to 1/2$ of $ U(\sigma) / \Gamma(1-\frac{1+l}{p})$ for $l=p-1$  when $p$ is equal to $1/2$  and more generally is a half-integer.
 
 We have  chosen a  value for $l$  in the set  $l=-1,-\frac{1}{2},0$, which are all the same mod $p$. Indeed to match the divergence of $U(\sigma)$, we have to take $l=-\frac{1}{2}$ which is
 equal to $p-1$, and $\Gamma(1- \frac{1+ l}{p}) = \Gamma(0)$. Taking this value $l=-\frac{1}{2}$, the continuation from integer $p$ to $p=\frac{1}{2}$ becomes smooth with finite intersection numbers. 
 Note also that  (\ref{onepoint}) is an exact form expression, while (\ref{exact2}) is an expansion.  This leads to the verification of the expansion (\ref{3/2u}) by the exact $U(\sigma)$ for $p=\frac{1}{2}$.

The result for $p=\frac{1}{2}$ in (\ref{onepoint})  may also be written as
\be
U(\sigma) = \sum_{g} \frac{c^{4g-2}3^{4g-2}\sigma^{6g-3}}{(4g-2)!}
\oint\frac{dz}{2i\pi} (1 + z)(1- \frac{z}{3})^{4g-2}\frac{1}{z^{g+1}}
\ee
where the contour integral, denoted by $C_g^{(0)}$, is
\be\label{Cg0}
 C_g^{(0)} =\oint \frac{dz}{2i\pi} (1 + z) (1 - \frac{z}{3})^{4g-2}\frac{1}{z^{g+1}} = \frac{(4g-2)!}{g! (3g-1)!}.
 \ee
 
 There are signs such as $(-1)^g$ in the expression. We absorb those minus signs by  a redefinition of positive
intersection numbers.
\be
<\tau_{6g-5-2l}>_g = \biggl( \frac{3^{3g-1}}{(3g)!}\biggr) \frac{1}{(g-1)!} = \frac{3^{3g-2}}{g! (3g-1)!}, \hskip 3mm (p = \frac{1}{2})
\ee
 The differences of three choices of $l=-1,-\frac{1}{2},0$ make just the shift of the notation of suffix of $\tau$, i.e. $6g-5-2l$, since they are same as mod $p$.
In the following, we take $l=-1$ value for R type punctures for $p=\frac{1}{2}$.
 \vskip 3mm
 \section{The two-point function for $p = \frac{1}{2}$}
 \vskip 2mm
  
 
  For the two-point connected function $U(\sigma_1,\sigma_2)$, the integral representation (\ref{s-point}) reads, after tuning of the source matrix,
 \ba\label{U2p2a}
 U(\sigma_1,\sigma_2) &=& 4 \oint \frac{dy_1 dy_2}{(2\pi i)^2} (y_1+ \frac{\sigma_1^2}{y_1^3})(y_2+ \frac{\sigma_2^2}{y_2^3})
  e^{c'\sigma_1(3 y_1-\frac{\sigma_1^2}{y_1^3}) + c' \sigma_2 (3 y_2- \frac{\sigma_2^2}{y_2^3})}\nonumber\\
 &\times& \frac{1}{(y_1^2- \frac{\sigma_1^2}{y_1^2} - y_2^2 + \frac{\sigma_2^2}{y_2^2})^2 + 4 ( \sigma_1+ \sigma_2)^2}
 \ea

We use the change of variables (\ref{zy}) for the two-point function (\ref{U2p2a}). We then rescale  $y_i$ by $y_i/\sigma_i$ and obtain
\ba\label{U2}
 U(\sigma_1,\sigma_2) &=& \frac{4\sigma_1^2}{\sigma_2^2} \oint \frac{dy_1 dy_2}{(2\pi i)^2} (y_1+ \frac{\sigma_1^6}{y_1^3})(y_2+ \frac{\sigma_2^6}{y_2^3})
  e^{c_1 (3 y_1-\frac{\sigma_1^6}{y_1^3}) + c_2  (3 y_2- \frac{\sigma_2^6}{y_2^3})}\nonumber\\
 &\times& \frac{1}{(y_1^2- \frac{\sigma_1^6}{y_1^2} - \frac{\sigma_1^2}{\sigma_2^2} y_2^2 + \frac{\sigma_1^2\sigma_2^4}{y_2^2})^2 + 4 \sigma_1^4 ( \sigma_1+ \sigma_2)^2}
 \ea
Expanding the above equation in powers of $\sigma_1^{n_1}\sigma_2^{n_2}$, we obtain as coefficients the intersection numbers $<\tau_{n_1}\tau_{n_2}>_g$ for genus $g$. The selection rule for $n_1$ and $n_2$ follows (\ref{RR}). For the  two-point function  $n_1+n_2= 6g$.

\vskip 2mm
 {\bf{3-1 Equation for $\sigma_1\sigma_2^{6g-1}, \sigma_1^3\sigma_2^{6g-3}$ and $\sigma_1^5\sigma_2^{6g-5}$ }}
 \vskip 2mm
 
From (\ref{U2}), it  follows immediately  that there is no  linear term in  $\sigma_1$  such as  $\sigma_1\sigma_2^{6g-1}$. Thus we have $<\tau_1\tau_{6g-1}>_g = 0$.
We also find that $\sigma_1^3\sigma_2^{6g-3}$ and $\sigma_1^5\sigma_2^{6g-5}$ are absent from (\ref{U2}) 
 and thus $<\tau_1\tau_{6g-3}>_g$ and $<\tau_5\tau_{6g-5}>_g $ also vanish.

   \vskip 2mm
 {\bf{3-2 Coefficient of $\sigma_1^2\sigma_2^{6g-2}$}}
 \vskip 2mm
A term of order $\sigma_1^2$ is present  as a  coefficient in a front of the integral  (\ref{U2}).   All other $\sigma_1$ being  set  to zero, 
 the integral for the term of order $\sigma_1^2$  reduces to
 \be
  U(\sigma_1,\sigma_2) = \frac{4\sigma_1^2}{\sigma_2^2} \oint \frac{dy_1 dy_2}{(2\pi i)^2} \frac{1}{y_1^3} (y_2+ \frac{\sigma_2^6}{y_2^3})
  e^{c_1 (3 y_1) + c_2  (3 y_2- \frac{\sigma_2^6}{y_2^3})}
 \ee
 The residue at the  pole  $y_1=0$ is computed after  expanding  $e^{3c_1y_1}$, and the integral over $y_2$ is found to be the same as for the one point function $<\tau_{6g-3}>$.
 This leads to an equation, which is similar to the integer-$p$ string equation, and thus we give it here the same name.
 
 \vskip 2mm
 {\bf{String equation}},
 The coefficient of $\sigma_1^2$ leads to the relation
  \be
  <\tau_2\tau_{6g-2}>_g =  \frac{3^2}{2} <\tau_{6g-3}>_g
  \ee
  \vskip 2mm
  This  is similar to the string equation for integer  $p$, ($6g-2\to 6g-3)$.  
 For the correspondence to  the usual Kontsevich model, we have the selection rule $6g - 6 + 6 = n_1+ n_2$ while $3g - 3 + 2 n'_1+ n'_2$, thus for $n_1+n_2 = 2(n'_1+n'_2) + 2$.
 By setting $n_1= 2, n'_1=0$, we obtain  $n_2= 2 n'_2$.
 Thus the above equation may be interpreted as $<\tau'_0\tau'_{3g-1}>$ giving the string equation of Kontsevich model $<\tau'_0\tau'_{3g-1}> = <\tau'_{3g-2}>$.

  \vskip 3mm
 {\bf{3-4 Coefficient of $\sigma_1^n\sigma_2^{6g-n}$}}
 \vskip 2mm

{$\bullet$} $<\tau_4\tau_8>$
\vskip 2mm
For $<\tau_4\tau_8>$ one  considers the terms of scale   $c^{8}\sigma^{12}$  (in general $c^{4g}\sigma^{6g}$, here with $g=2$). The power $c^8$ consists of the  non-vanishing  terms $c_1^2c_2^6$ and $c_1^4c_2^4$.

From (\ref{U2}), neglecting higher powers in  $\sigma_1$ than   $\sigma_1^4$ ,   the integral reduces to
\be\label{48a}
I = \frac{\sigma_1^2}{\sigma_2^2} \oint \frac{dy_1 dy_2}{(2\pi i)^2} \frac{e^{3c_1y_1}}{y_1^3}(y_2+ \frac{\sigma_2^6}{y_2^3})e^{  c_2  (3 y_2- \frac{\sigma_2^6}{y_2^3})}
\frac{1}{(1- \frac{\sigma_1^2 y_2^2}{\sigma_2^2 y_1^2} + \frac{\sigma_1^2\sigma_2^4}{y_1^2 y_2^2})^2}
\ee
The term $\sigma_1^4\sigma_2^8$ is obtained as the residue at the  $y_1=0$ pole, through the expansion of the denominator,
\ba
I &=& \biggl(\frac{3^4 c_1^4}{4!}\biggr)  \frac{2\sigma_1^4}{\sigma_2^4} \oint  \frac{ dy_2}{(2\pi i)}y_2^3 e^{  c_2  (3 y_2- \frac{\sigma_2^6}{y_2^3})}\nonumber\\
&=& \frac{3^5}{16}c_1^4c_2^4 \sigma_1^4\sigma_2^8
\ea

 \vskip 2mm
{$\bullet$} $<\tau_4\tau_{14}>$
\vskip 2mm
The term $\sigma_1^4$ is the  same in this case, so we use (\ref{48a}), and up to order $\sigma_2^{14}$ we may write this as
\be\label{414}
I = \biggl(\frac{\sigma_1^4 c_1^4 3^4}{4!}\biggr) \frac{2}{\sigma_2^4}\oint \frac{dy_1 dy_2}{(2\pi i)^2} \frac{1}{y_2} (y_2^4-\frac{\sigma_2^{12}}{y_2^4} ) e^{  c_2  (3 y_2- \frac{\sigma_2^6}{y_2^3})}
\ee

From this, we obtain the coefficient of $\sigma_1^4\sigma_2^{14}$,
\be
I =\frac{3^8}{2\cdot 7!} c_1^4c_2^8 \sigma_1^4 \sigma_2^{14}
\ee

  \vskip 2mm
{$\bullet$} $<\tau_4\tau_{20}>$
\vskip 2mm
The integral (\ref{414})   for $<\tau_4\tau_{6g-4}>$ gives for $g=4$ 

 \be
 <\tau_4\tau_{20}> = \frac{3^{12}}{16\cdot 10!} c_1^4c_2^{12}
 \ee
   \vskip 2mm
  
  For general genus $g$,
 \ba\label{46g-4}
 <\tau_4\tau_{6g-4}> &=& 
 \frac{3^{3g}}{2 g (g-2)! (3g-2)!} c_1^4 c_2^{4g-4}\nonumber\\
 &&= \frac{(g-1)(3g-1)}{3^2}{2}<\tau_{6g-3}>
 \ea
This formula results from  from (\ref{U2p2a}) ;  taking  the pole at $y_1=\infty$  for $c_1^2 \sigma_1^4$,  the term of order $c_2^{4g-2}\sigma_2^{6g-4}$  is
expressed as
\be
I= \frac{3^{4g}}{2 \cdot 2!(4g-2)!}\oint \frac{dz}{2i\pi} (1-z)(1+z)(1- \frac{z}{3})^{4g-2} \frac{1}{z^{g+1}}
\ee
The intersection number $<\tau_4\tau_{6g-4}>$ corresponds to the $<\tau'_1\tau'_{3g-2}>$ of Kontsevich model, which suggests that  it is related to a dilaton equation.

\vskip 2mm
{$\bullet$} $<\tau_6\tau_{6g-6}>$
\vskip 2mm
The coefficient  of $\sigma_1^6$ is obtained by neglecting higher order terms in (\ref{U2}),
\ba
I &=& \biggl(\frac{\sigma_1^2}{\sigma_2^2}\biggr) \oint \frac{dy_1 dy_2}{(2\pi i)^2} \frac{1}{y_1^3} (y_2+ \frac{\sigma_2^6}{y_2^3})
  e^{c_1 (3 y_1) + c_2  (3 y_2- \frac{\sigma_2^6}{y_2^3})}\nonumber\\
&\times&( 1 + \frac{2 \sigma_1^2 y_2^2}{\sigma_2^2 y_1^2} - \frac{2 \sigma_1^2\sigma_2^4}{y_1^2y_2^2}- \frac{6\sigma_1^4\sigma_2^2}{y_1^4}  -\frac{\sigma_1^4y_2^4}{\sigma_2^4y_1^4} - \frac{\sigma_1^4\sigma_2^8}{y_1^4y_2^4}) 
\ea
For general $g$, we obtain
\be
<\tau_6\tau_{6g-6}> =   \frac{4g-5}{20\cdot g! (3g-4)!} 3^{3g-1} c_1^6 c_2^{6g-6}
\ee
This formula is obtained, first by taking the pole at   $y_1=\infty$  in (\ref{U2p2a}), which gives a $\sigma_1^6$ term ; then  the coefficient of t $\sigma_2^{6g-6}$ is given by
\be
I = \oint \frac{dz}{2i\pi} (3 - 10 z + 3 z^2)(1+ z) (1- \frac{z}{3})^{4g-6}\frac{1}{z^{g+1}}.
\ee

 \vskip 2mm
 {\section {Three-point function function for $p= \frac{1}{2}$}}
 \vskip 3mm
The  three-point function $U(\sigma_1,\sigma_2,\sigma_3)$  is given  by a triple integral involving a $3\times3$ determinant. The connected part is given by  the longest cycles of the determinant, and it is  given by an integral over $y_1,y_2$ and $y_3$ :
 \ba\label{yyy3}
&& U_c(\sigma_1,\sigma_2,\sigma_3) = - {\it i}\oint \frac{dy_1dy_2dy_3}{(2i \pi)^3} (\frac{\sigma_1^2}{y_1^3} + y_1) (\frac{\sigma_2^2}{y_2^3} + y_2) (\frac{\sigma_3^2}{y_3^3} + y_3)
 \nonumber\\
 &&\times \frac{1}{(2 {\it i}(\sigma_1 + \sigma_2) 
    -  \frac{\sigma_1^2}{y_1^2} + y_1^2 + \frac{\sigma_2^2}{y_2^2} - y_2^2) (2 {\it i}(\sigma_2 + \sigma_3) - \frac{\sigma_2^2}{y_2^2} + 
     y_2^2 + \frac{\sigma_3^2}{y_3^2} - y_3^2)} \nonumber\\
     && \times \frac{1}{(2 {\it i} (\sigma_1 + \sigma_3) + \frac{\sigma_1^2}{y_1^2} - y_1^2 - \frac{\sigma_3^2}{
     y_3^2} + y_3^2)}e^{\sum_{i=1}^3 c_i \sigma_i (3 y_i - \frac{\sigma_i^2}{y_i^3})}
 \ea
 or, after the rescaling $y_i\to y_i/\sigma_i$, by
  \ba\label{yyy3}
&& U(\sigma_1,\sigma_2,\sigma_3)
    = \frac{1}{(\sigma_1\sigma_2\sigma_3)^2}\oint \frac{dy_1dy_2dy_3}{(2i \pi)^3} (\frac{\sigma_1^6}{y_1^3} + y_1) (\frac{\sigma_2^6}{y_2^3} + y_2) (\frac{\sigma_3^6}{y_3^3} + y_3)
 \nonumber\\
 &&\times \frac{1}{(2 {\it i}(\sigma_1 + \sigma_2) - 
     \frac{\sigma_1^{4}}{y_1^2} + \frac{y_1^2}{\sigma_1^2} + \frac{\sigma_2^{4
     }}{y_2^2} -\frac{ y_2^2}{\sigma_2^2}) } \frac{1}{(2 {\it i}(\sigma_2 + \sigma_3) - 
     \frac{\sigma_2^{4}}{y_2^2} + \frac{y_2^2}{\sigma_2^2} + \frac{\sigma_3^{4
     }}{y_3^2} -\frac{ y_3^2}{\sigma_3^2}) }\nonumber\\
     &&
     \frac{1}{(2 {\it i}(\sigma_1 + \sigma_3) + 
     \frac{\sigma_1^{4}}{y_1^2} - \frac{y_1^2}{\sigma_1^2} - \frac{\sigma_3^{4
     }}{y_3^2} + \frac{ y_3^2}{\sigma_3^2}) }
      e^{c_1  (3 y_1 - \frac{ \sigma_1^6}{y_1^3})+ c_2  (3 y_2 - \frac{ \sigma_2^6}{y_2^3})+ c_3  (3 y_3 - \frac{ \sigma_3^6}{y_3^3})
      }\nonumber\\
 \ea
  
   In the  three-point case, the condition (\ref{RR}) reads
  \be
 2g + 1 = \frac{1}{3}(n_1+n_2+n_3)
 \ee
 where $n_i$  denotes the index of $\tau_{n_i}$, $n_i\ge 1$.
 
 For genus $g=0$,  the only possible choice is $<\tau_1^3>$, but the contour integral leads to
 \be
 <\tau_1^3>_{g=0}=0
 \ee
       \vskip 3mm
  {\bf 5-1 Equations for $\tau_1$}. Since there is no term linear  in $\sigma_1$  in (\ref{yyy3}), we have
    $
  <\tau_1\tau_{n_2}\tau_{6g+2-n_2}> = 0
  $
  \vskip 2mm
  
 \vskip 2mm
{\bf{5-2 Equation for $\tau_2$}}
\vskip 2mm
From (\ref{yyy3}), the term of order $\sigma_1^2$ is obtained by
\ba\label{yyy/2}
I &=& - \frac{\sigma_1^2}{(\sigma_2\sigma_3)^2} \oint \frac{dy_1dy_2dy_3}{(2i \pi)^3} \frac{e^{3 c_1 y_1}}{y_1^3}(\frac{\sigma_2^6}{y_2^3} + y_2) (\frac{\sigma_3^6}{y_3^3} + y_3)
 e^{ c_2  (3 y_2 - \frac{ \sigma_2^6}{y_2^3})+ c_3  (3 y_3 - \frac{ \sigma_3^6}{y_3^3})}\nonumber\\
 &\times&\frac{1}{(2 {\it i}(\sigma_2 + \sigma_3) - 
     \frac{\sigma_2^{4}}{y_2^2} + \frac{y_2^2}{\sigma_2^2} + \frac{\sigma_3^{4
     }}{y_3^2} -\frac{ y_3^2}{\sigma_3^2}) }
 \ea
  
  The denominator is expanded as
  \be\label{3to2}
  \frac{\sigma_2^2}{y_2^2} + \frac{\sigma_2^4}{y_2^4}\biggl(\frac{y_3^2 }{\sigma_3^2} - \frac{\sigma_3^4}{y_3^2} - 2i (\sigma_2+\sigma_3)\biggr)
  + O(\sigma_2^6)
 \ee
  
 The first term, i.e.  $ \frac{\sigma_2^2}{y_2^2}$,  gives  a non-vanishing residue in $\oint \frac{dy_2}{2i\pi} \frac{1}{y_2}= 1$, and the dependence in $\sigma_2$ is canceled by the factor in the front of (\ref{yyy/2}).
 The integral over $y_3$ gives (\ref{onepoint}), i.e. $<\tau_{6g-3}>\sigma_3^{6g-2}$. Thus from the first term, we obtain $<\tau_2\tau_0\tau_4>= \frac{(3c_1)^2}{2}<\tau_{6g-3}> \sigma_1^2\sigma_3^{6g-2}$,
 which  reduces to the two-point function $U(\sigma_1,\sigma_3)$, with the selection rule of $6g= n_1+n_3$.  
 This $\tau_0$ plays the role of what was the dilaton equation in the case of  integer $p$.

 The second term gives the coefficient of $\sigma_2^2$,
 \ba
 &&\frac{\sigma_1^2\sigma_2^2}{\sigma_3^4}\oint \frac{dy_2}{2i\pi}\frac{e^{3c_2y_2}}{y_2^3}\oint \frac{dy_3}{2i\pi} \frac{1}{y_3}(y_3^2- \frac{\sigma_3^{12}}{y_3^6})e^{c_3(3y_3-\frac{\sigma_3^6}{y_3^3})}\nonumber\\
 &&- \frac{2 i \sigma_1^2\sigma_2^2}{\sigma_3}\oint \frac{dy_2}{2i\pi}\frac{e^{3c_2y_2}}{y_2^3}\oint \frac{dy_3}{2i\pi}(y_3+ \frac{\sigma_3^6}{y_3}) e^{c_3(3y_3-\frac{\sigma_3^6}{y_3^3})}
 \ea
 There is no $(\sigma_1\sigma_2\sigma_3)^2$ term, which  is consistent with  the selection rule $6g + 3= n_1+ n_2 + n_3$. The second term of the above equation gives $\sigma_1^2\sigma_2^2\sigma_3^5$, i.e, $<\tau_2\tau_2\tau_5>_{g=1}$.

 We have
   \be
  <\tau_2\tau_2\tau_{6g-1}> = <\tau_2\tau_{6g-2}>= <\tau_{6g-3}>
  \ee
  which can be interpreted hereto as a string equation.
  
  Noting  that  two point function  (\ref{U2}) is expressed as
  \ba\label{U2b}
 &&U(\sigma_2,\sigma_3) = \frac{4\sigma_2^2}{\sigma_3^2} \oint \frac{dy_2 dy_3}{(2\pi i)^2} (y_2+ \frac{\sigma_2^6}{y_2^3})(y_3+ \frac{\sigma_3^6}{y_3^3})
  e^{c_2 (3 y_2-\frac{\sigma_2^6}{y_2^3}) + c_3  (3 y_3- \frac{\sigma_3^6}{y_3^3})}\nonumber\\
 &\times& \frac{1}{(y_2^2- \frac{\sigma_2^6}{y_2^2} - \frac{\sigma_2^2}{\sigma_3^2} y_3^2 + \frac{\sigma_2^2\sigma_3^4}{y_3^2})^2 + 4 \sigma_2^4 ( \sigma_2+ \sigma_3)^2}\nonumber\\
 &&= \frac{ i}{\sigma_2^2\sigma_3^2(\sigma_2+\sigma_3)} \oint \frac{dy_2 dy_3}{(2\pi i)^2} (y_2+ \frac{\sigma_1^6}{y_2^3})(y_3+ \frac{\sigma_2^6}{y_3^3})
 e^{c_2 (3 y_2-\frac{\sigma_1^6}{y_2^3}) + c_3  (3 y_3- \frac{\sigma_2^6}{y_3^3})}\nonumber\\
&&\biggl( \frac{1}{\frac{y_2^2}{\sigma_2^2}- \frac{\sigma_2^4}{y_2^2}-\frac{y_3^2}{\sigma_3^2}+ \frac{\sigma_3^4}{y_3^2}- 2 i (\sigma_2+\sigma_3)}
- \frac{1}{\frac{y_2^2}{\sigma_2^2}- \frac{\sigma_2^4}{y_2^2}-\frac{y_3^2}{\sigma_3^2}+ \frac{\sigma_3^4}{y_3^2} +  2 i (\sigma_2+\sigma_3)}\biggr)\nonumber\\
 \ea
and comparing with (\ref{yyy/2}), we obtain the relation
\be
<\tau_2\tau_{n_2}\tau_{n_3}> = <\tau_{n_2-1}\tau_{n_3}> + <\tau_{n_2}\tau_{n_3-1}>
\ee
 The above equation may be interpreted as a puncture (string) equation. 
  For $g=1$, we have $<\tau_2\tau_3\tau_4>$ from (\ref{yyy/2}), and it also satisfies the string equation.

 \vskip 3mm
 {\section {Correlation functions for $p= \frac{3}{2}$}}
 \vskip 3mm

The spin $p=\frac{3}{2}$ has  marked points of both NS ($l=0,-\frac{1}{2}$) and R ($l=-1$)  types.  Or we take $l= 0,-\frac{1}{2}, \frac{1}{2}$ due to mod $p$. The case $l=\frac{1}{2}$ is equivalent to $l=-1$ (mod $p$).
 As shown in \cite{BrezinHikami2020}, the relevant integrand for $p=\frac{3}{2}$ has the factor
\be
g(y) = e^{\sum_{i=1}^s c_i\sigma_1 (5 y_i^3-\frac{10\sigma_1^2}{y_i} + \frac{\sigma_1^4}{y_i^5} ) }
\ee
after the rescaling $y_i\to y_i/\sqrt{\sigma_i}$. The selection rule  for an  $s$-point function of R-type is here
($p=\frac{3}{2}$),
\be\label{selection}
\frac{10}{3}(g-1) +  s = \sum_{i=1}^s n_i + \frac{2}{3}\sum_{i=1}^s l_i
\ee
When  all marked points belong to the R-type, i.e.  $l_i=-1$ or $l_i= \frac{1}{2}$,
the selection rule (\ref{selection}) reads for $l_i= -1$,
\be
6g- 6 + 3s = \frac{9}{5}\sum_{i=1}^s n_i
\ee
and for the choice  $l_i= \frac{1}{2}$,
\be
6g - 6 + \frac{6}{5}s = \frac{9}{5}\sum_{i=1}^s n_i
\ee

We have for the one-point function of R-type, 
\ba\label{onepoint3/2}
U(\sigma) &=& \frac{1}{\sigma}\oint \frac{dy}{2i \pi} (y+ \frac{\sigma^2}{y^3}) e^{c' \sigma (5 y^3-\frac{10\sigma^2}{y} + \frac{\sigma^4}{y^5} ) }\nonumber\\
&=&5 c'^2 \sigma^5 - \frac{7\cdot 5^4}{48} c'^6 \sigma^{15} + \frac{79\cdot11\cdot 5^5}{7\cdot 2^7\cdot 3^2} c'^{10}\sigma^{25} + \cdots
\ea
From these coefficients, we have  $<\tau_n> = <\tau_{\frac{10g-5}{3}}>, (g=2,5,8,...)$.
\be\label{tau5}
<\tau_5>_{g=2}= 5 c'^2,\hskip 3mm <\tau_{15}>_{g=5} =  \frac{7\cdot 5^4}{48} c'^6, \hskip 3mm <\tau_{25}>_{g=8} =  \frac{79\cdot11\cdot 5^5}{7\cdot 2^7\cdot 3^2} c'^{10}
\ee

Since there is a rescaling between $y$ and $\sigma$, these integrals are reduced to the contour integral over $z= \sigma^2$,
\be
I = \frac{1}{(\frac{4g-2}{3})!}\oint \frac{dz}{2i\pi} \frac{(1+z)(5 - 10 z + z^2)^{\frac{4g-2}{3}}}{z^{g+1}}
\ee
which gives for  $g=8$, $I= \frac{79\cdot11\cdot 5^5}{7\cdot 2^7\cdot 3^2}$, and for $g=11$, $I= -\frac{191\cdot 5^7\cdot 13^2}{11\cdot 7\cdot 3^4\cdot 2^{10}}$.

To verify that the  continuation to $p=\frac{3}{2}$ makes sense, we compare the results (\ref{tau5}) with the  previous integer-$p$  intersection numbers.
The results for the coefficients of $\sigma^{15}$ and $\sigma^{25}$  correspond to $g=5$ and $g=8$. These cases were computed in \cite{BrezinHikami2} for general $p$. For $g=5$, it reads
\ba\label{g5}
<\tau_{n,l}>_{g=5} &=& (p-1)(p-3)(p-9)(1+ 2p)(3+4p)(32 p^4-162 p^3 + p^2 \nonumber\\
&+& 326 p + 163)
 \frac{1}{p^4}\frac{1}{11! 4^5 3}\frac{\Gamma(1- \frac{9}{p})}{\Gamma(1- \frac{1+l}{p})}
\ea
To compare with the second term of (\ref{tau5}), we multiply  it by $1/(4p^5)=(\frac{2}{3})^5/4$. We take $c' = c/40$. 
This agrees with (\ref{g5}) if one takes into account the factor $1/5!$, which comes from the regularization of the  infinite factor $\Gamma(1-\frac{9}{p})= \frac{1}{5!}
\Gamma(0)$ for $p=3/2$. Note that the divergence $\Gamma(0)$  is cancelled by the denominator gamma factor $\Gamma(1- \frac{1+l}{p})$ if we take  $l= p-1= \frac{1}{2}$, which originated from (\ref{def}).
(Note this cancellation is $\Gamma(0)/\Gamma(0)=1$, and it is not the cancellation in the limit of $\epsilon\to 0$ for $p=\frac{3}{2}+ \epsilon$ since $\epsilon\to 0$ gives some additional numerical facor.)

For the third term of (\ref{onepoint3/2}) of order $\sigma^{25}$, we confirm the  continuation to half-integer $p$  with the expression for the intersection number $<\tau_{n,m}>_{g=8}$ in \cite{BrezinHikami2}.
Thus the  continuation to $p=\frac{3}{2}$ for the computation of  intersection numbers of $p$-spin curves,  is valid as for $p=\frac{1}{2}$, through the contour integral calculation.

For $p=\frac{3}{2}$, there are also intersection numbers from NS punctures ($l\ne -1$). 
 Putting $p=\frac{3}{2}$ in (\ref{onepoint3/2}), we obtain the intersection numbers for NS punctures.
The components of the spin $p=\frac{3}{2}$ are $l=-1,-\frac{1}{2}$ and 0.  The intersection number  $<\tau_{n,l}>$ is given by the coefficient of $\sigma^{n+\frac{2}{3}(1+l)}$. We have  the intersection numbers  $<\tau_{1,0}>_{g=1}$, $<\tau_{8,-\frac{1}{2}}>_{g=3}$, $<\tau_{11,0}>_{g=4}$. Therefore they belong to the NS type punctures ($l\ne -1$).

These results may  also  be derived from  (\ref{zy}). We have
\be\label{onepoint3/2a}
U(\sigma) = \frac{1}{\sigma} \oint \frac{dy}{2i\pi} (y+ \frac{\sigma^2}{y^3}) e^{c' \sigma (5 y^3 - \frac{10 \sigma^2}{y} + \frac{\sigma^4}{y^5})}
\ee
Keeping the exponential term $e^{c \sigma (5 y^3 )}$,  the other terms are expanded for small $\sigma$. Changing  variable  to $y= \sigma^{-\frac{1}{3}}x$, 
we have
\be\label{onepoint3/2b}
U(\sigma) = \frac{1}{\sigma^{\frac{5}{3}}}\oint \frac{dx}{2i\pi} ( x + \frac{\sigma^{\frac{10}{3}}}{x^3}) e^{5 c' x^3 - \frac{10 c' \sigma^{\frac{10}{3}}}{x} 
+ \frac{c' \sigma^{\frac{20}{3}}}{x^5}}
\ee
This integral is computed  in a   small $\sigma$ expansion   
\be
U(\sigma)= \frac{1}{6}(5c')^{\frac{2}{3}}(-1)^{-\frac{2}{3}}\Gamma(\frac{1}{3}) \sigma^{\frac{5}{3}} + 5 c'^2 \sigma^5 + O(\sigma^{\frac{25}{3}})
\ee
where the integral yields gamma functions for this NS type  as in (\ref{3/2u}).  The third term, of order $\sigma^{\frac{25}{3}}$,  is evaluated by an expansion of (\ref{onepoint3/2b}),
\be\label{3/2expansion}
I =  \sigma^{\frac{25}{3}} \int \frac{dz}{2i\pi} e^{5c' x^3} ( -\frac{10 c'^2}{x^5} - \frac{1000 c'^3}{6x^2} + \frac{c'}{x^8} + \frac{50 c'^2}{x^5})
\ee
By the change of variable $x= (5c')^{-\frac{1}{3}}z$, this term , after integration from $-\infty$ to 0 for $z$,
\be
I = -\frac{29}{27} \Gamma(-\frac{7}{3})\cdot \frac{25}{3}(5c')^{\frac{1}{3}}c'^3 \sigma^{\frac{25}{3}}
\ee
This result is consistent with the last term  of (\ref{3/2u}). Thus for NS punctures, we confirm the continuation from integer $p$ to the half-integral case,  within again an infinite renormalization proportional to  $\Gamma(0)$.
The expansion of (\ref{3/2expansion}), yields  a series of terms $\sigma^{5/3}$, $\sigma^5$, $\sigma^{25/3}$ of the form $\sigma^{n+ \frac{2}{3}(1+ l)}$, with
$l= 0$, $l=-1$ (or 1/2  mod $p= \frac{3}{2}$), and $l= -\frac{1}{2}$. Thus these terms provide the results for  $<\tau_{1,0}>$, $<\tau_{4,\frac{1}{2}}>$ and $<\tau_{8,-\frac{1}{2}}>$, respectively.

\vskip 2mm
{\bf{$\bullet$ Two-point function for $p=\frac{3}{2}$}}
\vskip 2mm
 For the two-point function $U(\sigma_1,\sigma_2)$ of $p=\frac{3}{2}$, we have
 \ba\label{U2p2}
 U(\sigma_1,\sigma_2) &=& 4 \oint \frac{dy_1 dy_2}{(2\pi i)^2} (y_1+ \frac{\sigma_1^2}{y_1^3})(y_2+ \frac{\sigma_2^2}{y_2^3})
 \frac{1}{(y_1^2- \frac{\sigma_1^2}{y_1^2} - y_2^2 + \frac{\sigma_2^2}{y_2^2})^2 + 4 ( \sigma_1+ \sigma_2)^2}\nonumber\\
 &\times&e^{c_1 \sigma_1 (5 y_1^3 - \frac{10 \sigma_1^2}{y_1} + \frac{\sigma_1^4}{y_1^5})+ c_1 \sigma_2 (5 y_2^3 - \frac{10 \sigma_2^2}{y_2} + \frac{\sigma_2^4}{y_2^5})}
 \ea
After rescaling, we obtain
\ba
&&U(\sigma_1,\sigma_2) = \frac{4}{(\sigma_1\sigma_2)^{2/3}} \oint \frac{dy_1 dy_2}{(2\pi i)^2} (y_1+ \frac{\sigma_1^{10/3}}{y_1^3})(y_2+ \frac{\sigma_2^{10/3}}{y_2^3})\nonumber\\
&&\times  \frac{1}{(y_1^2 \sigma_1^{-2/3}- \frac{\sigma_1^{8/3}}{y_1^2} - y_2^2 \sigma_2^{-2/3}+ \frac{\sigma_2^{8/3}}{y_2^2})^2 + 4 ( \sigma_1+ \sigma_2)^2}\nonumber\\
&&\times e^{c_1 (5 y_1^3 - \frac{10 \sigma_1^{10/3}}{y_1} + \frac{\sigma_1^{20/3}}{y_1^5})+ c_2  (5 y_2^3 - \frac{10 \sigma_2^{10/3}}{y_2} + \frac{\sigma_2^{20/3}}{y_2^5})}
\nonumber\\
 \ea
From this expression, we  obtain the terms which contribute to  $<\tau_{n_1,l_1}\tau_{n_2,l_2}> \sigma_1^{n_1+ \frac{2}{3}(1+ l_1)}\sigma_2^{n_2+ \frac{2}{3}(1+l_2)}$.

For the two-point function,
we have $n_1+n_2= \frac{10}{3}g$, and  the R-type is easily evaluated 
\be
<\tau_2\tau_8>_{g=3}  = \frac{11\cdot 5^3}{4} c_1^2c_2^2,\hskip 3mm <\tau_3\tau_7>_{g=3} = 500 c_1^2c_2^2,\hskip 3mm <\tau_4\tau_6>_{g=3} = \frac{5^4}{2}c_1^2c_2^2
\ee
and $<\tau_1\tau_9>= <\tau_5\tau_5>=0$. 
 
The term $\sigma_1^{2/3}$  gives $<\tau_{0,0}\tau_{n_2,l_2}>$,
\be
I = \frac{\sigma_1^{2/3}}{\sigma_2^{2/3}}\oint \frac{dy_1dy_2}{(2i \pi)^2} \frac{y_1 (y_2+ \frac{\sigma_2^{10/3}}{y_2^3})}{y_1^4} e^{c_1 (5 y_1^3 )} e^{c_2  (5 y_2^3 - \frac{10 \sigma_2^{10/3}}{y_2} + \frac{\sigma_2^{20/3}}{y_2^5})}
\ee
The  integrals over $y_1$ and $y_2$ factorize. The coefficient of   $\sigma^{\frac{2}{3}} (= \sigma^{n+\frac{2}{3}(1+l)})$ is $<\tau_{0,0}>$.
The integral over $y_1$ gives a  factor  $\Gamma(-\frac{2}{3})$. The integral over $y_2$, differs from (\ref{onepoint3/2})  by a factor $\frac{1}{\sigma_2}$.
Thus we have a string equation in the  $p=\frac{3}{2}$ case for an NS-type  puncture,  similar to  what we found for integer-$p$ ,
\be\label{stringE}
<\tau_{0,0}\tau_{n_2,l_2}>_g = <\tau_{n_2-1,l_2}>_g
\ee
An example is $<\tau_{0,0}\tau_{2,0}>_{g=1} = <\tau_{1,0}>_{g=1}$, where $\tau_{0,0}$ and $\tau_{2,0}$ are both NS punctures $(l=0)$. For the mixture of NS and R types, we have
$<\tau_{0,0}\tau_{6,-1}>_{g=2} = <\tau_{5,-1}>_{g=2}$. The term  $\tau_{6,-1}$ is obtained from the coefficient  of $\sigma_2^{6}$ as a contour integral 
around $y_2=0$. 
 
 Next we obtain the   $\sigma_1^{4/3}$-term , which corresponds to $<\tau_{1,-\frac{1}{2}}>$. 
 \ba
 I &=& \frac{4\sigma_1^{4/3}}{\sigma_2^{4/3}}\oint \frac{dy_1dy_2}{(2i\pi)^2} \frac{1}{y_1^5} (y_2^2+ \frac{\sigma_2^{10/3}}{y_2^2})(y_2^2- \frac{\sigma_2^{10/3}}{y_2^2})
 \nonumber\\
 && e^{c_1 (5 y_1^3 )} e^{c_2  (5 y_2^3 - \frac{10 \sigma_2^{10/3}}{y_2} + \frac{\sigma_2^{20/3}}{y_2^5})}
 \ea
Thie coefficient of $\sigma_1^{4/3}$ gives $<\tau_{1,-\frac{1}{2}}>$ since $\sigma_1^{4/3} = \sigma_1^{n_1+ \frac{2}{3}(1+ l_1)}$ with $n_1=1$ and $l_1= -\frac{1}{2}$ , and above $I$ gives $<\tau_{1,-\frac{1}{2}}\tau_{n_2,l_2}>$.  In this case the selection rule reads $(g-1)\frac{10}{3} + \frac{4}{3} = n_2 + \frac{2}{3}l_2$.

We have seen the string equation ($L_{-1}$)  related to $<\tau_{0,0}>$ in (\ref{stringE}). The dilaton equation ($L_0$)  is related to $<\tau_{1,0}>$. The coefficient here-above $<\tau_{1,-\frac{1}{2}>}$ is related to
$L_{-\frac{1}{2}}$, where $L_n = z^{n}\partial_z$.

\vskip 2mm
{\bf{$\bullet$ Three-point function for $p=\frac{3}{2}$}}
\vskip 2mm

 \ba\label{yyy}
&& U(\sigma_1,\sigma_2,\sigma_3) 
    = \frac{1}{(\sigma_1\sigma_2\sigma_3)^{2/3}}\oint \frac{dy_1dy_2dy_3}{(2i \pi)^3} (\frac{\sigma_1^{10/3}}{y_1^3} + y_1) (\frac{\sigma_2^{10/3}}{y_2^3} + y_2) (\frac{\sigma_3^{10/3}}{y_3^3} + y_3)
 \nonumber\\
 &&\times \frac{1}{(2 {\it i}(\sigma_1 + \sigma_2) - 
     \frac{\sigma_1^{8/3}}{y_1^2} + \sigma_1^{-2/3}y_1^2 + \frac{\sigma_2^{8/3
     }}{y_2^2} -\sigma_2^{-2/3} y_2^2) }\nonumber\\
     &&\frac{1}{(2 {\it i}(\sigma_2 + \sigma_3) - \frac{\sigma_2^{8/3}}{y_2^2} + 
     \sigma_2^{-2/3}y_2^2 + \frac{\sigma_3^{8/3}}{y_3^2} - \sigma_3^{-2/3}y_3^2)} \nonumber\\
     && \times \frac{1}{(2 {\it i} (\sigma_1 + \sigma_3) + \frac{\sigma_1^{8/3}}{y_1^2} - \sigma_1^{-2/3}y_1^2 - \frac{\sigma_3^{8/3}}{
     y_3^2} + \sigma_3^{-2/3}y_3^2)}\nonumber\\
      &\times&e^{c_1  (5 y_1^3 - \frac{10 \sigma_1^{10/3}}{y_1} + \frac{\sigma_1^{20/3}}{y_1^5})+ c_2  (5 y_2^3 - \frac{10 \sigma_2^{10/3}}{y_2} + \frac{\sigma_2^{20/3}}{y_2^5}) +
      c_3  (5 y_3^3 - \frac{10 \sigma_3^{10/3}}{y_3} + \frac{\sigma_3^{20/3}}{y_3^5})}\nonumber\\
 \ea
where we have  rescaled  $y_i\to \sigma_i^{-1/3}y_i$.
 
 For $g=0$, we have $<\tau_{0,0}\tau_{0,0}\tau_{0,-\frac{1}{2}}>\sigma_1^{\frac{2}{3}}\sigma_2^{\frac{2}{3}}\sigma_3^{\frac{1}{3}}$, in which each $ \sigma$ is of the form $\sigma^{n+ \frac{2}{3}(1+ l)}$. The factor $\sigma_3^{1/3}$ comes from the expansion of $(\sigma_2+\sigma_3)$ in the denominator. It satisfies the selection rule  of $p=\frac{3}{2}$ for $<\tau_{n_1,l_1}\tau_{n_2,l_2}\tau_{n_3,l_3}>$,
 \be\label{selec}
 (g-1)\frac{10}{3} + 3 = \sum n_i + \frac{2}{3}\sum l_i
 \ee
 This is consistent with the  three-point function for an integer $p$-spin curve in  (A.8)  \cite{BrezinHikami03},
 \be\label{333}
 <\tau_{0,l_1}\tau_{0,l_2}\tau_{0,l_3}>_{g=0} = \delta_{l_1+l_2+l_3,p-2}
 \ee
 This  result for integer $p$,   is continued to the half-integer $p=\frac{3}{2}$. This equation follows from, 
 \be
 0 = \sum_{i=1}^3 n_i - (1- \frac{2}{p}) + \frac{1}{p}\sum_{i=1}^3 l_i
 \ee
 with  $g=0$, $n_i=0$.
There are other cases for $g=0$, which satisfy the selection rule (\ref{selec}). They correspond to  mixtures of the NS and R types.
\ba\label{5}
&&<\tau_{0,-1}\tau_{0,-1}\tau_{1,0}>, \hskip 3mm <\tau_{1,-\frac{1}{2}}\tau_{0,-\frac{1}{2}}\tau_{0,-1}>,\hskip 3mm <\tau_{0,-\frac{1}{2}}\tau_{0,-\frac{1}{2}}\tau_{1,-1}>,
\nonumber\\
&& <\tau_{1,-1}\tau_{0,0}\tau_{0,-1}>
\ea
The existence of such terms can be checked from the expansion of (\ref{yyy}). We find a term  $\sigma_1^{\frac{4}{3}}\sigma_3^{\frac{1}{3}}$ from the expansion of $(\sigma_1+\sigma_3)\sigma_1^{\frac{2}{3}}$. And $\sigma_2^0$ is obtained as in the $p=\frac{1}{2}$ case in (\ref{3to2}). Thus we obtain
$<\tau_{1,-\frac{1}{2}}\tau_{0,-\frac{1}{2}}\tau_{0,-1}>$ for $g=0$.
This correlation  is $\sigma_1^{\frac{4}{3}}\sigma_2^{\frac{1}{3}}\sigma_3^{0}$.  This $\tau_{0,-1}$ plays the role of a dilaton equation, as was the case for $p=\frac{1}{2}$. This may be  expressed conveniently by the introduction of a  field $\phi_i$,  conjugate to  $\sigma^{\frac{i}{3}}$, i.e. $<\tau_{n+ 
\frac{2}{3}(1+ l)}>
= <\phi_{3n+ 2(1+l)}>$.
Then this  correlation is expressed from the primary field $\phi_i$ as $<\phi_0\phi_1\phi_4>$. The correlation  $<\tau_{0,0}\tau_{0,0}\tau_{0,-\frac{1}{2}}>$ in (\ref{333}) is expressed as 
$<\phi_1\phi_2\phi_2>$.
Thus we found that there are two non-vanishing correlations for $<\phi_1\phi_1\phi_2>$ and $<\phi_4\phi_1\phi_0>$.

The fields  $\phi_0$ and $\phi_3$ re related to R-type punctures.
These terms give the structure factor $C_{ijk}$ as $\phi_i\phi_j = \sum {C_{ij}}^k \phi_k$.
The structure constant $C_{ijk}$ satisfies WDVV equation \cite{Witten4,DVV},
\be
{C_{ij}}^m C_{mkl} = {C_{ik}}^m C_{m j l}
\ee
where $\phi_i\phi_j = \sum {C_{ij}}^k \phi_k$. The fields $\phi_i$ ($i=0,1,..,4$) are  primary fields ($g$=0) conjugate to  $\sigma^{n+ \frac{2}{3}(1+l)}$ ($\phi_i= \sigma^{\frac{i}{3}}$) and they own a ring structure.  We have $C_{014}$ and $C_{112}$ from the three-point functions. 
$C_{014}= <\phi_0\phi_1\phi_4>= <\tau_{0,-1}\tau_{0,-\frac{1}{2}}\tau_{1,-\frac{1}{2}}>_{g=0}$.

\vskip 2mm
\vskip 2mm
{\bf{$\bullet$ Four-point function for $p=\frac{3}{2}$}}
\vskip 2mm
A representation of the four-point functions for $p$ spin curves follows from the same  random matrix theory \cite{BrezinHikami03}. 
For genus $g=0$, we have
three different cases, which satisfy the selection rule for $p=\frac{3}{2}$, $\frac{2}{3}= \sum_{i=1}^4 n_i + \frac{2}{3}\sum_{i=1}^4 l_i$;
\ba\label{4point}
&&<\tau_{1,-\frac{1}{2}}\tau_{0,-\frac{1}{2}}\tau_{0,-\frac{1}{2}}\tau_{1,-\frac{1}{2}}>_{g=0}\sigma_1^{\frac{4}{3}}\sigma_2^{\frac{1}{3}}\sigma_3^{\frac{1}{3}}\sigma_4^{\frac{4}{3}},\nonumber\\
&&<\tau_{1,-\frac{1}{2}}\tau_{0,-\frac{1}{2}}\tau_{0,0}\tau_{1,-1}>_{g=0}\sigma_1^{\frac{4}{3}}\sigma_2^{\frac{1}{3}}\sigma_3^{\frac{2}{3}}\sigma_4,\nonumber\\
&&<\tau_{1,- 1}\tau_{0,0}\tau_{0,0}\tau_{1,- 1}>_{g=0}\sigma_1\sigma_2^{\frac{2}{3}}\sigma_3^{\frac{2}{3}}\sigma_4
\ea
These are decomposed as three-point functions  $<\tau_{1,-1}\tau_{0,0}\tau_{0,-1}>$ and $<\tau_{1,-\frac{1}{2}}\tau_{0,-\frac{1}{2}}\tau_{0,-1}>$, in a conformal block decomposition.

The four-point function is obtained from the longest cycles in the determinant  which occurs in the integral representation \cite{BrezinHikami03}. Neglecting  terms of higher genus, we write one of the cycles (another longest cycle is obtained by permutation)  as
\ba\label{4pointa}
&&U(\sigma_1,\sigma_2, \sigma_3,\sigma_4) = \frac{1}{(\sigma_1\sigma_2\sigma_3\sigma_4)^{2/3}}\oint \prod \frac{dy_i}{2i\pi} (\prod_{i=1}^4 y_i) e^{\sum c_i 5 y_i^3}\nonumber\\
&& \times \frac{1}{(\frac{y_1^2}{\sigma_1^{2/3}}- \frac{y_2^2}{\sigma_2^{2/3}} + 2 i (\sigma_1+\sigma_2))}\frac{1}{(\frac{y_2^2}{\sigma_2^{2/3}}- \frac{y_3^2}{\sigma_3^{2/3}} + 2 i (\sigma_2+\sigma_3))}\nonumber\\
&& \times \frac{1}{(\frac{y_3^2}{\sigma_3^{2/3}}- \frac{y_4^2}{\sigma_4^{2/3}} + 2 i (\sigma_3+\sigma_4))}\frac{1}{(\frac{y_4^2}{\sigma_4^{2/3}}- \frac{y_1^2}{\sigma_1^{2/3}} + 2 i (\sigma_1+\sigma_4))}
\ea
From this expression, we obtain $\sigma_1^{4/3}\sigma_2^{1/3}\sigma_3^{4/3}\sigma_4^{1/3}$ term by  expansion, which gives 
$<\tau_{1,-\frac{1}{2}}\tau_{0,-\frac{1}{2}}\tau_{0,-\frac{1}{2}}\tau_{1,-\frac{1}{2}}>$ in (\ref{4point}). The  other two terms in (\ref{4point}) are not derived from (\ref{4pointa}).

The structure constant $C_{ijk}$ satisfies WDVV equation \cite{Witten4,DVV},
\be
{C_{ij}}^m C_{mkl} = {C_{ik}}^m C_{m j l}
\ee
where $\phi_i\phi_j = \sum {C_{ij}}^k \phi_k$. The fields $\phi_i$ ($i=0,1,..,4$) are primary fields ($g$=0) conjugate to $\sigma^{n+ \frac{2}{3}(1+l)}$ ($\phi_i= \sigma^{\frac{i}{3}}$) and they  make a ring structure.  We have $C_{014}$  from the three-point functions 
$C_{014}= <\phi_0\phi_1\phi_4>= <\tau_{0,-1}\tau_{0,-\frac{1}{2}}\tau_{1,-\frac{1}{2}}>_{g=0}$. The four point function of $g=0$ (\ref{4point}) are
${C_{14}}^0C_{014}$. The field $\phi_0$ connects the  two conformal blocks $C_{ij0}$, and $\phi_0$ is an R type puncture. So, ifor genus 0, two
R type punctures $\phi_0$ are contracted in pair in a four-point functions for $p=\frac{3}{2}$ . This paired contraction seems  to be a characteristic property of R-type punctures $\phi_0$.

From the relation to the generating function $F$ as $C_{ijk}= \frac{\partial^3 F}{\partial t_i\partial t_j\partial t_k}$ \cite{Witten1,DVV, BrezinHikami03}, we have for $p=\frac{3}{2}$,
\be\label{F3/2}
F = \frac{1}{2}t_{0,0}^2 t_{0,-\frac{1}{2}} + t_{1,-\frac{1}{2}}t_{0,-\frac{1}{2}}t_{0,-1} + \frac{1}{4}  t_{1,-\frac{1}{2}}^2 t_{0,-\frac{1}{2}}^2 + O(t^5)
\ee
The first term is NS, and the second term involves an R-type puncture. For  integer $p$, the corresponding $F$ is $p=3$, as shown in \cite{BrezinHikami03}. The first term
is the same as for $p=3$, but $p=3$ does not have the R-type as the second term of (\ref{F3/2}).

 \vskip 3mm
 \section{Negative integer spins : $p=-2$ and $p=-3$}
\vskip 2mm
 Whereas for positive integer $p$, there are no R-punctures, they do appear  when $p$ is continued to negative integer values. Such R- punctures
 must then occur pairwise as implied by the selection rule (\ref{RR}) for $p=-2$.
 \vskip 2mm
{\bf{$\bullet$  $p= - 2$}}
\vskip 2mm
When $p= -2$, the one-point correlation function reads
\be\label{strong}
U(\sigma) = \frac{i}{2} \oint \frac{dy}{2i\pi} (y+ \frac{1}{y^3}) e^{c' \frac{1}{\sigma} y^4 \frac{1}{(1+ y^4)^2}}
\ee
where $c' = 16 c$.

After the rescaling $y \to \sigma^{\frac{1}{4}} y$, 
\be\label{y}
U(\sigma) = \frac{i}{2}\sigma^{\frac{1}{2}}\oint \frac{dy}{2i \pi} (y + \frac{1}{\sigma y^{3}}) e^{c'  y^4 \frac{1}{(1 + \sigma y^4)^2}}
\ee
The expansion in powers of  $\sigma$ is the same as the expansion found in the unitary matrix model. In \cite{BrezinHikami20}, we have computed $U(\sigma)$ from the integral of (\ref{uintegral}), and confirmed the equivalence with a unitary matrix model. The expansion in powers of $y$ from (\ref{y}) gives the same result. The integral reduces to a Gaussian integral,  with a power series in $\sigma$ with coefficients involving $\Gamma(\frac{1}{2})$.
Expanding for small $\sigma$, we find terms  $\sigma^{n- \frac{1}{2}}$, which  are of the form $\sigma^{n+\frac{1}{p}(1+ l)}$ with $l=0$.
Those terms belong thus to the NS-type \cite{BrezinHikami2020}.

From (\ref{y}), we have also lower genus terms. The term of order $\sigma^{-\frac{1}{2}}$ is $\sigma^{n-\frac{1}{2}(1+l)}$ with $n=0,l=0$, which is the genus 0 term $<\tau_{0,0}>_{g=0}$.
The next order is  $\sigma^{\frac{1}{2}}$, which is $n=1,l=0$. This term is genus 1, and gives $<\tau_{1,0}>_{g=1}$.  
Note that in the expression of $U(\sigma)$ in (\ref{3/2u}), there is no divergent gamma  factors for $p=-2$.
\vskip 2mm
{\bf{$\bullet$ Two-point function for $p= - 2$}}
\vskip 2mm
 For the two-point function $U(\sigma_1,\sigma_2)$ for $p= - 2$, we have after  scaling, 
 \ba\label{U2p2}
 U(\sigma_1,\sigma_2) &=& (\sigma_1\sigma_2)^{\frac{1}{2}} \oint \frac{dy_1 dy_2}{(2\pi i)^2} (y_1+ \frac{1}{\sigma_1 y_1^3})(y_2+ \frac{1}{\sigma_2 y_2^3})
 e^{c_1'  y_1^4\frac{1}{(1+\sigma_1 y_1^4)^2}+ c_2'   y_2^4\frac{1}{(1+\sigma_2 y_2^4)^2}}\nonumber\\
&& \frac{1}{(\sigma_1^{1/2}y_1^2- \frac{1}{\sigma_1^{1/2}y_1^2} -\sigma_2^{1/2} y_2^2 + \frac{1}{\sigma_2^{1/2}y_2^2})^2 + 4 ( \sigma_1+ \sigma_2)^2}\nonumber\\
 \ea

Selecting the term proportional to  $\sigma_1^{1/2}$, one gets
\be
U(\sigma_1,\sigma_2) = \sigma_1^{1/2}\sigma_2^{1/2}\int \frac{dy_1}{2i\pi} y_1 e^{c_1' y_1^4} \int \frac{dy_2}{2i\pi} (y_2+ \frac{1}{\sigma_2 y_2^3}) e^{c_2' y_2^4 \frac{1}{(1+ \sigma_2 y_2^4)^2}} + O(\sigma_1)
\ee
This term provides the result for $<\tau_{1,0}\tau_{n_2,0}> \sigma_1^{1/2}\sigma_2^{n_2- \frac{1}{2}}$. Since the above integral  is factorized, one finds that $<\tau_{1,0}\tau_{n_2,0}> \sim
<\tau_{n_2,0}>$. This corresponds to a dilaton equation. In the unitary matrix model, it is known that there is no string equation but there is a dilaton equation \cite{Gross}.

There is also   a term proportional  to $\sigma_1\sigma_2$ in $U(\sigma_1,\sigma_2)$, which corresponds to 
$\sigma_1^{n_1-\frac{1}{2}(1+ l_1)}\sigma_2^{n_2-\frac{1}{2}(1+l_2)}$ with $l_1=-1,l_2=-1$. This is a pair of R-type marked points $(l_i=-1)$.
This paired R-type is expressed as $<\tau_{1,-1}\tau_{1,-1}> \sigma_1\sigma_2$, which corresponds to  a genus $g=2$ term.
\vskip 2mm
{\bf{$\bullet$ Three-point function for $p= - 2$}}
\vskip 2mm
 \ba\label{yyy}
U(\sigma_1,\sigma_2,\sigma_3)&=&(\sigma_1\sigma_2\sigma_3)^{1/2} \oint \frac{dy_1dy_2dy_3}{(2i \pi)^3} (\frac{1}{\sigma_1 y_1^3} + y_1) (\frac{1}{\sigma_2 y_2^3} + y_2) (\frac{1}{\sigma_3 y_3^3} + y_3)
 \nonumber\\
 &&\times \frac{1}{(2 {\it i}(\sigma_1 + \sigma_2) - 
     \frac{1}{\sigma_1^{1/2}y_1^2} + \sigma_1^{1/2}y_1^2 + \frac{1}{\sigma_2^{1/2}y_2^2} -\sigma_2^{1/2} y_2^2) }\nonumber\\
   &&\times \frac{1}{(2 {\it i}(\sigma_2 + \sigma_3) - 
     \frac{1}{\sigma_2^{1/2}y_2^2} + \sigma_2^{1/2}y_2^2 + \frac{1}{\sigma_3^{1/2}y_3^2} -\sigma_3^{1/2} y_3^2) }\nonumber\\
   &&\times \frac{1}{(2 {\it i}(\sigma_1 + \sigma_3) - 
     \frac{1}{\sigma_3^{1/2}y_3^2} + \sigma_3^{1/2}y_3^2 + \frac{1}{\sigma_1^{1/2}y_1^2} -\sigma_1^{1/2} y_1^2) }\nonumber\\
      &\times&e^{c_1 y_1^4  \frac{1}{(1+ \sigma_1 y_1^4)^2}+ c_2 y_2^4  \frac{1}{(1+ \sigma_2 y_2^4)^2} + c_3 y_3^4  \frac{1}{(1+ \sigma_3 y_3^4)^2}}\nonumber\\
 \ea
 Let us note that there is a term of the form $<\tau_{1,0}\tau_{0,-1}\tau_{0,-1}>\sigma_1^\frac{1}{2}\sigma_2^0\sigma_3^0$. This is two R-type marked points and one NS-type, which is genus $g=0$.
 
There is also a term  $\sigma_1^{3/2}\sigma_2\sigma_3$ , after expansion of the denominator of (\ref{yyy}), similar to the two-point function, and
the coefficient is $<\tau_{2,0}\tau_{1,-1}\tau_{1,-1}>$, which is a paired of R-type marked points plus one NS-type  with $g=3$.
\vskip 2mm
{\bf{$\bullet$ Pairwise punctures of R-type }}
\vskip 2mm
The selection rule  for  negative $p$  is
\be\label{RRR}
3g -3 + s = \sum_{i=1}^s n_i + (g-1)(1 - \frac{2}{p}) + \frac{1}{p} \sum_{i=1}^s l_i
\ee
For $p=-2$, the spin components $l_i$ take the values $l_i= 0$ or $l_i=-1$. The number $s$  of  marked points is divided as $s= m_R+ m_{NS}$, where $m_R$ is the number with $l_i=-1$
(R type), and $m_{NS}$ is the number with $l_i=0$ (NS type). We set the number of boundary
holes to zero.
From (\ref{RRR}), we have for $p=-2$ 
\be
g- 1 + \frac{1}{2}m_{R} + m_{NS} = \sum_{i=1}^s n_i 
\ee
Since $g,m_{NS},n_i$ are all integers, we have the condition that 
\be\label{even}
m_R = 2 N
\ee
where $N$ is a natural number, namely $m_R$ is even.
Thus Ramond punctures should be paired, if $m_R\ne 0$ .

\vskip 2mm
{\bf{$\bullet$ strong coupling expansion for $p=-2$}}
\vskip 2mm
It is known that there is a phase transition in the unitary matrix model between  weak coupling (small $\sigma$) and  strong coupling (large $\sigma$) regions
\cite{GrossWitten,BrezinGross}.
From (\ref{strong}), we expand the one-point function $U(\sigma)$ for  large $\sigma$, after the replacement  $y= x^{1/2}$,
\ba\label{unitary}
U(\sigma) &=& \frac{1}{2} \int \frac{dx}{2\pi} (1 + \frac{1}{x^2}) \sum \frac{1}{n!} (\frac{c'}{\sigma})^n \frac{x^{2n}}{(1 + x^2)^{2n}}\nonumber\\
&=& \frac{1}{2}\sum \frac{1}{n!} (\frac{c'}{\sigma})^n \frac{\Gamma(n- \frac{1}{2})\Gamma(n+ \frac{1}{2})}{\Gamma(2n)}
\ea
This expansion agrees with the known expansion for large $\sigma$, i.e. the  strong coupling expansion \cite{BrezinHikami20} and with the unitary matrix model \cite{GrossWitten,BrezinGross}. The expression of the one-point function of (\ref{unitary}) in the strong coupling region provides a confirmation that the $p=-2$ model is
equivalent to the unitary matrix model. 

 We did not include here
  the logarithmic term, considered earlier  in \cite{BrezinHikami20}. If  included, the expression would be modified by an additional coefficient $N$ of the logarithmic term. 
The above expression provides  the result for $N=0$.
 
This model of $p=-2$ has an interesting transition between  the weak coupling regime and the strong coupling regime, as in lattice QCD at large N  \cite{BrezinGross}.

\vskip 2mm
{\bf{$\bullet$  $p= - 3$}}
\vskip 2mm
For $p=-3$, the one-point function reads
\ba\label{-3p}
U(\sigma) &=& \frac{i}{2} \oint \frac{dy}{2i \pi} (y + \frac{1}{y^3})e^{-16 c \frac{1}{\sigma^2} (8i)\frac{y^{10}- y^6}{(1+ y^4)^4}}\nonumber\\
&=& \frac{i}{2}\sigma^{2/3}\oint \frac{dy}{2i \pi} (y + \frac{1}{\sigma^{4/3} y^3}) e^{c' (y^6-\sigma^{4/3}y^{10})/((1+ \sigma^{4/3}y^4)^4}
\ea
where we have used the scaling $y \to \sigma^{1/3}y$.
We have terms of order $\sigma^{n-\frac{1}{3}(1+ l)}$ with $ l=-1,0,1$. 

For the two point function $U(\sigma_1,\sigma_2)$ of $p= - 3$, we have, after similar re-scaling,
 \ba\label{U2p2}
 U(\sigma_1,\sigma_2) &=& (\sigma_1\sigma_2)^{\frac{2}{3}} \oint \frac{dy_1 dy_2}{(2\pi i)^2} (y_1+ \frac{1}{\sigma_1^{4/3} y_1^3})(y_2+ \frac{1}{\sigma_2^{4/3} y_2^3})\nonumber\\
 && e^{c' (y_1^6-\sigma_1^{4/3}y_1^{10})/((1+ \sigma_1^{4/3}y_1^4)^4 + c' (y_2^6-\sigma_2^{4/3}y_2^{10})/((1+ \sigma_2^{4/3}y_2^4)^4 }
 \nonumber\\
&& \frac{1}{(\sigma_1^{2/3}y_1^2- \frac{1}{\sigma_1^{2/3}y_1^2} -\sigma_2^{2/3} y_2^2 + \frac{1}{\sigma_2^{2/3}y_2^2})^2 + 4 ( \sigma_1+ \sigma_2)^2}
 \ea
  The expansion  of $U(\sigma_1,\sigma_2)$ gives for instance a 
 $(\sigma_1\sigma_2)^{2}$ term
which is
a pair of R-type $<\tau_{2,-1}\tau_{2,-1}>_{g=3}$.  

\vskip 2mm
{\bf{$\bullet$ strong coupling expansion for $p=-3$}}
\vskip 2mm
There is a phase transition between a weak coupling  and a strong coupling region,  similar to
$p=-2$.
The strong coupling power series in $1/\sigma$ expansion is  obtained from the first equation of (\ref{-3p}). 
\ba
U(\sigma) &=& \frac{1}{2}\sum (\frac{c'}{\sigma^2})^n \frac{1}{n!} \int \frac{dy}{2\pi} \frac{y^{6n-1}(y^4-1)^n}{(1+ y^4)^{4n-1}}\nonumber\\
&=& \frac{1}{16\pi} \sum (\frac{c'}{\sigma^2})^n \frac{1}{n!} \int_0^\infty dt t^{\frac{3}{2}n - 1} \frac{(t-1)^n}{(1+ t)^{4n-1}}\nonumber\\
&=&\frac{1}{16\pi} \sum_{n=1}^\infty (\frac{c'}{\sigma^2})^n  \sum_{m=0}^n \frac{1}{m!(n-m)!}(-1)^m \frac{\Gamma(\frac{5}{2}n-m)\Gamma(\frac{3}{2}n+ m-1)}{\Gamma(4n-1)}
\nonumber\\
\ea
In the case $p=-2$, we find that the strong coupling expansion agrees with character expansions \cite{BrezinHikami20}. For $p=-3$, we don't know the corresponding character expansions.
\vskip 3mm

 \section{Random supermatrices}

In \cite{BrezinHikami2020}, we have discussed a supermatrix formulation, which gives the correlation functions for two $p$ and $p'$ mixed spin-curves. Both spin values of  $p$ and $p'$ are tuned by the
two external sources which represent the bosonic and fermionic parts.
We use the explicit representation Eq.(113) of \cite{BrezinHikami2020}, which reads 
 \ba\label{ppoint}
&&U(\sigma_1,....,\sigma_n) = <{\rm str}e^{\sigma_1 M} \cdots {\rm str}e^{\sigma_n M} >\nonumber\\
&=& \oint \prod_{i=1}^n \frac{du_i}{2i \pi} e^{c_1 \sum_{i=1}^n [(u_i+\frac{\sigma_i}{2})^{p+1}-(u_i - \frac{\sigma_i}{2})^{p+1}]}
 e^{c_2 \sum_{i=1}^n [(u_i+\frac{\sigma_i}{2})^{p'+1}-(u_i - \frac{\sigma_i}{2})^{p'+1}]}\nonumber\\
&\times& {\rm det}\frac{1}{u_i-u_j + \frac{1}{2}(\sigma_i+\sigma_j)}
\ea
with $c_1= \frac{n}{p^2-1}\sum_{i=1}^{p-1} \frac{1}{r_i^{p+1}}$, and  $c_2= \frac{m}{p'^2-1}\sum_{i=1}^{p'-1} \frac{1}{\rho_i^{p'+1}}$. 
In this formula, we have a freedom to chose a  set of  values for $p$ and $p'$. 
\vskip 2mm
\vskip 2mm
{\bf{$\bullet$ $D_n$ singularity}}
\vskip 2mm

The choice of $p=n+1$ and $p'= -2$ is related to a
singularity of $D_n$ type, with a Landau-Ginzburg (LG) potential  $W = x^{n-1} + xz^2 - 2\lambda z$. There are generalizations to KP hierarchies, including positive and negative
power series  in the Landau-Ginzburg potential \cite{DVV, Hanany}.
This LG potential becomes
$W= x^{n-1}-\lambda^2/x$ after  integration over $z$. Other choices of $p$ and $p'$ may give other singularities and topologies
for Riemann surfaces. In the previous sections, we have discussed  half-integer and negative integer $p$-spin cases. The supermatrix formulation \cite{BrezinHikami2020}, makes it possible to extend
the spin moduli space to a space with two different spins $p$ and $p'$.

If we specialize  (\ref{ppoint})  to $p'= -2$, arbitrary $p$, it yields a $D_{p+1}$ singularity, ($x^2+ y^2 z + z^{p}=0,$ $p\ge 3$), .
We then use the variable $y$ of (\ref{zy}), to write
\be
g(y) = e^{c_1' \sigma^{p+1} \frac{1}{y^{2p+2}}((y^2-i)^{2p+2}-(y^2-i)^{2p+2}) + c_2' \frac{1}{\sigma} \frac{y^4}{(1+ y^4)^2}}
\ee
For the case $p=3$ ($D_4$), this gives
\be
g(y) = e^{c_1' \sigma^4 (-16 i )(y^6 -7 y^2 + 7 \frac{1}{y^2} - \frac{1}{y^6}) + c_2' \frac{1}{\sigma} \frac{y^4}{(1+ y^4)^2}}
\ee
After rescaling $y\to \sigma^{-2/3}y$, and absorbing a factor $(-16i)$ into $c_1'$, we obtain
\be
g(y) = e^{c_1' (y^6 - 7 \sigma^{8/3} y^2 + 7 \sigma^{16/3} y^{-2} - \sigma^8 y^{-6}) + c_2' \sigma^{5/3} y^{-4}(1+ \sigma^{8/3}/y^4)^{-2}}
\ee
The one-point function becomes
\be
U(\sigma) = \frac{1}{\sigma^{4/3}} \int \frac{dy}{2i \pi} (y+ \frac{\sigma^{8/3}}{y^3})  e^{c_1' (y^6 - 7 \sigma^{8/3} y^2 + 7 \sigma^{16/3} y^{-2} - \sigma^8 y^{-6}) + c_2' \sigma^{5/3} y^{-4}(1+ \sigma^{8/3}/y^4)^{-2}}
\ee 
If $c_2'=0$, there is no R-term. The contour integrals around $y=0$ for the terms $\sigma^n$ ($n$ is integer) vanish .   But if $c_1',c_2\ne 0$, there are  terms of R-type, at  order  $\sigma^n$. For instance, expanding the exponent for small $\sigma$, we obtain at lowest order,
\be
I = (c_1'(c_2')^2 \frac{1}{2}\sigma^2 - 6 c_1'c_2' \sigma^3) \oint \frac{dy}{2i\pi} (\frac{1}{y})
\ee
These integer powers of $\sigma$ come from $\sigma^{n+ \frac{1}{3}(1+ l)}$ with $l=-1$.

As discussed in \cite{BrezinHikami2020},  for the supermatrices with $p' < 0$, logarithmic terms appear, which correspond  to open boundaries of Riemann surfaces.
For the unitary matrix case ($p'=-2$), this logarithmic term appears with  a  coefficient $N$, which is  the number of D-branes \cite{BrezinHikami20}. 
\vskip 2mm
{\bf{$\bullet$ continuation to half-integers with boundaries, as in   $p=\frac{1}{2}$  }}
\vskip 2mm

We consider $p=\frac{1}{2}$ with a logarithmic potential, since for $p=\frac{1}{2}$ there are  only R- punctures in the bulk. The addition of boundaries by a logarithmic potential may provide additional 
R-punctures on the  boundary.

We thus consider the correlation function $U(\sigma_1,...\sigma_s)$  for $p=\frac{1}{2}$ and $p'=-1$ in (\ref{ppoint}).  The one-point function reads
\be
U(\sigma)= \frac{1}{2}\oint \frac{du}{2i\pi} e^{c_1 ( \frac{\sigma}{2})^{3/2} [(u+ 1)^{3/2} - (u-1)^{3/2}] + N {\rm log}[(u+1)/(u-1)]}
\ee
From (\ref{zy}), $u= \frac{i}{2}(y^2-\frac{1}{y^2})$,
\ba\label{oriented}
&&U(\sigma) = \frac{i}{2}\oint \frac{dy}{2i\pi} (y+ \frac{1}{y^3}) e^{c_1 (\frac{\sigma}{2})^{3/2} (\frac{i}{2})^{1/2} (3y - \frac{1}{y^3})+ 2N {\rm log}[(y^2-i)/(y^2+i)]}\nonumber\\
&&=\frac{i}{2\sigma}\oint \frac{dy}{2i\pi} (y+ \frac{\sigma^2}{y^3}) e^{c_1' \sigma (3y - \frac{\sigma^2}{y^3})+ 2N {\rm log}[(1-i\sigma/y^2)/(1+i\sigma/y^2)]}\nonumber\\
&&= \frac{i}{2\sigma}\oint \frac{dy}{2i\pi} (y+ \frac{\sigma^2}{y^3})\biggl( 1 - \frac{4i N \sigma}{y^2} - \frac{8N^2 \sigma^2}{y^4} + 
\frac{4i}{3}(N+ 8N^3)\frac{\sigma^3}{y^6} \nonumber\\
&&+ \frac{16}{3}(N^2+ 2 N^4) \frac{\sigma^4}{y^8} - \frac{i}{5}(12 N+ 160N^3+ 128 N^5) \frac{\sigma^5}{y^{10}}
+ \cdots \biggr) e^{c_1' \sigma (3y - \frac{\sigma^2}{y^3})}\nonumber\\
\ea

where the second line is obtained  after scaling $y\to \sigma^{-1/2}y$, and the last line comes from the expansion in powers of the logarithmic term.

The expansion for small $\sigma$ provides the  power series,
\ba\label{p1/2log}
U(\sigma) &=& \frac{i}{2}  [ - 4 i N - \frac{3}{2}(1- 24 N^2) c_1'^2 \sigma^3 + 9 i (N + 4 N^3) c_1'^4 \sigma^6\nonumber\\
 &&- \frac{27}{80}(1 - 40 N^2 -32 N^4) c_1'^6 \sigma^9 + \cdots ]
 \ea

Thus we have  found  that the logarithmic term yields additional factors in the expression (\ref{onepoint}) of $U(\sigma)$.  
 
 Let us quote the result for the one-point function of  integer
$p$-th curve from \cite{BrezinHikami05}.
\be\label{Unm}
U(\sigma) = \frac{1}{\sigma}\oint \frac{du}{2i \pi} e^{-\frac{1}{p+1} [ (u + \frac{\sigma}{2})^{p+1} - (u- \frac{\sigma}{2})^{p+1}]+ N{\rm log}(u + \frac{\sigma}{2})
/ (u- \frac{\sigma}{2})}
\ee
By choosing the integration path around the cut, with $x=\sigma u^p$, it becomes
\ba\label{pN}
&&U(\sigma) = \frac{1}{p\sigma^{1+ 1/p}\pi} \int_0^\infty dx x^{1/p-1} e^{-x} e^{-\frac{p(p-1)}{3! 4} \sigma^{2+ 2/p} x^{1 - 2/p} + \cdots}\nonumber\\
&&\times [ 1 + N (\sigma^{1+1/p}x^{-1/p} + \frac{1}{12}\sigma^{3+3/p}x^{-3/p} + \cdots) ]+ \cdots\nonumber\\
&&= - (\frac{p-1}{24} + \frac{N^2}{2})\frac{1}{\pi} \sigma^{1 + \frac{1}{p}} \Gamma(1 - \frac{1}{p}) +- (\frac{p}{24}N + \frac{1}{12}N^3) \frac{1}{\pi} \sigma^{2+ \frac{2}{p}}\Gamma(1 - \frac{2}{p})\nonumber\\
&&- \frac{1}{144}[ \frac{(p-1)(p-3)(1+ 2p)}{40} + (3p+1) N^2 + 2 N^4] \frac{1}{\pi}\sigma^{3 + \frac{3}{p}} \Gamma(1 - \frac{3}{p}) + \cdots\nonumber\\
\ea
This expression is a modification of (\ref{3/2u}) with the boundary logarithmic potential.

It is remarkable that (\ref{p1/2log}) agrees with the above expression (\ref{pN}) except for the gamma-function factor, which is  divergent for $p=\frac{1}{2}$. This agreement of continuation to $p=\frac{1}{2}$ has been discussed in section 2 without logarithmic term. Here we check the continuation to half integer $p$ from integer $p$ in the presence of the logarithmic potential, except $\Gamma(0)$.
Note that   the powers of $ \sigma$ and polynomials in $N$ agree between (\ref{p1/2log}) and (\ref{pN}) with $p=\frac{1}{2}$.

In \cite{BrezinHikami05}, we have obtained   the intersection numbers
\be\label{polynomial}
<\tau_{1,0}>= \frac{p-1 + 12 N^2}{24},\hskip 3mm<\tau_{2,1}> = \frac{1}{24}(pN+ 2 N^3)
 \ee
Thus  $<\tau_3>$ of $p=\frac{1}{2}$ in the second term of (\ref{p1/2log}) corresponds to $<\tau_{1,0}>$ in (\ref{polynomial}).

  \vskip 2mm
    \section{Summary  and discussion}
    \vskip 3mm
      In this article, we have investigated the correlation functions of   half-integral   and  negative integer-spins marked points by   a random matrix formulation. We have shown how to evaluate the correlation functions in  power series of $\sigma$ for  R-type punctures, which provide integer powers of $\sigma$.  The half-integral spin  $p$ cases include Ramond marked point $(l=-1)$ mod $p$.   We  have confirmed the selection rule (\ref{RR}) for half integer spin $p$ in $s$ point functions.
 It is remarkable that the selection rule established  for positive integer $p$  (\ref{RR}) remains valid for half integer and for negative integers. 
  
   For the negative $p=-2$, we found that R type punctures should be  paired in (\ref{even}).  We  have shown  pair-wise punctures in the three=point functions in (\ref{yyy}). 
    We have investigated the correlation functions of such R-type marked points  in  details for the case of $p=\frac{1}{2}$. We have  found that the structures between these intersection numbers are described by the equations, which have  a similarity with the Virasoro algebra of integer $p$. 
   
  For genus zero and $p=\frac{3}{2}$, we have calculated  the three-point function, which  gives the structure constants $C_{ijk}$  for the  algebra of the primary fields $\phi_i$.  The characteristic new feature is the existence of $\tau_{0,-1}$ of R-type in a three-point function.   
  This R-type puncture is similar to the twisted field $\phi_{*}$, which appears  in $D_n$ singularity theory \cite{DVV}. We have explicitly evaluated the  three point correlation function and the structure constant $C_{ijk}$ for the primary field (genus 0). This structure extends  to other half-integral spin cases such as $p= \frac{5}{2}$.   
   In section seven, through a supermatrix formulation, we have discussed a $D_4$ singularity, which is related to R-type punctures \cite{Fan}.
  
     The half-integral spin $p$ may have interesting applications  to WZW models, which could be related to the fractional quantum Hall effect with half integral level $k=p-2$.
  Ramond marked points may have some application to the novel topological insulators and superconductors with Majorana fermion  excitations at the boundaries, since both R punctures and  Majorana fermion appear  pair-wise.

  
   \vskip 3mm 
 { \bf Acknowledgement}
 \vskip 2mm
 We are thankful to  Edward Witten for the discussions about Ramond pair-wise punctures. S.H. is supported by JSPS KAKENHI 19H01813. 

  \vskip 5mm

 \end{document}